\documentclass[10pt,twocolumn,english]{revtex4}
\usepackage[T1]{fontenc}
\usepackage[latin9]{inputenc}
\usepackage{amsmath}
\usepackage{graphicx}
\usepackage{amssymb}

\makeatletter
\@ifundefined{textcolor}{}
{%
 \definecolor{BLACK}{gray}{0}
 \definecolor{WHITE}{gray}{1}
 \definecolor{RED}{rgb}{1,0,0}
 \definecolor{GREEN}{rgb}{0,1,0}
 \definecolor{BLUE}{rgb}{0,0,1}
 \definecolor{CYAN}{cmyk}{1,0,0,0}
 \definecolor{MAGENTA}{cmyk}{0,1,0,0}
 \definecolor{YELLOW}{cmyk}{0,0,1,0}
 }

\makeatother

\usepackage{babel}

\begin{document}

\title{Laplacian growth in self-consistent Laplacian field : Effect of the
long-range interparticle interactions on the fractal dimension of
structures formed by their aggregation-limited diffusion.}

\author{F. Carlier, E. Brion, and V. M. Akulin}

\address{Laboratoire Aim\'{e} Cotton, B\^{a}t. 505, CNRS II, Campus d'Orsay,
Orsay Cedex F-91405, France}
\begin{abstract}
We numerically simulate the dynamics of aggregation of interacting
atomic clusters deposited on a surface. We show that the shape of
the structures resulting from their aggregation-limited random walk
is affected by the presence of a binary interparticle Laplacian potential
due to, for instance, the surface stress field. We characterize the
morphologies we obtain by their Hausdorff fractal dimension as well
as the so-called external fractal dimension, which appears more sensitive
to the potential. We demonstrate the relevance of our model by comparing
it to previously published experimental results for antymony and silver
clusters deposited onto graphite surface. 
\end{abstract}


\maketitle

\section{Free and biased diffusion models}

Highly ramified dendritic structures often appear in nature in such
different situations as cell colony growth \cite{matsu}, dielectric
breakdown \cite{niemeyer}, viscous fingering in Hele-Show cells \cite{kadanoff},
liquid crystal growth \cite{chiutti}, electro-deposition \cite{myochain2002},
single-atom \cite{hwang,howells} and atomic cluster depositions \cite{yoon,bardottiprl,bardottiss,brechi}.
Though the objects observed are not mathematically rigorous fractals,
the self-similarity they show on a certain range of the probe length
justifies the (strictly speaking improper) use of this term to qualify
them \cite{mandelbrot}.

The description common to all these systems relies on the diffusion
equation \begin{equation}
\partial_{t}P\left(t,\vec{r}\right)=D\triangle P\left(t,\vec{r}\right)\label{EQ1}\end{equation}
 written for the probability distribution $P\left(t,\vec{r}\right)$
and completed by the conditions imposed on the gradient $\overrightarrow{\nabla}P\left(t,\vec{r}\in\partial S\right)$
at the boundaries $\partial S$ of the area $S$ occupied by the structure
at a given moment of time $t$. In the quasistatic regime, when a
typical time of boundary variations always remains much longer than
$SD^{-1}$, the probability satisfy the Laplace equation $\triangle P\left(t,\vec{r}\right)=0$
with the boundary conditions \cite{batchelor} evolving as the structure
grows. This regime allows an elegant description based on conformal
mapping, which accounts for the structure dynamics \cite{schraiman,kadanoff}.

An alternative description relies on random walk models, including
the widely known Diffusion Limited Aggregation (DLA) model \cite{witten},
which yields a good visual agreement with experimentally observed
morphologies \cite{bardottiprl}. The models are based on a probabilistic
dynamical process\begin{equation}
\overrightarrow{r}_{i}(t)\overset{p\left(\overrightarrow{r},\overrightarrow{r}^{\prime}\right)}{\rightarrow}\overrightarrow{r}_{i}^{\prime}(t+\Delta t)\label{EQ2}\end{equation}
according to which the $i$-th particle located at a point $\overrightarrow{r}_{i}$
jumps into the point $\overrightarrow{r}_{i}^{\prime}$ within the
step $t\rightarrow\left(t+\Delta t\right)$ with the probability $p\left(\overrightarrow{r}_{i},\overrightarrow{r}_{i}^{\prime}\right)=p\left(\overrightarrow{r}_{i}^{\prime},\overrightarrow{r}_{i}\right)$.
This process is assumed to last until the particle touches another
particle and stops, thus contributing to an immobile aggregated dendrite.
A link between those two approaches was identified in \cite{kadanofflink},
giving rise to hybrid methods such as the iteration of random conformal
mappings \cite{procaccia,hastings,stepanov} which generate DLA-type
objects.

Still the results of experiments with atomic clusters deposited at
a surface and aggregated to dendritic structures in the course of
difussion, are just in a qualitative agreement with the DLA simulations
\cite{msdla}, whereas careful inspection reveals discrepancies in
compactness of the experimental and the DLA-induced structure, thus
indicating that some physical ingredients are missing in the model.
In the present work, we assume that the missing element is a Laplacian
field induced by the aggregated clusters and acting onto the moving
ones. The Laplacian field can be of different origins, and one of
the options is a surface stress field, which results from the mismatch
of the cristalline structures of the cluster and the substrate matherial
and yields mutual attraction of the deposited clusters. In this new
setting, the actual motion of clusters is no longer a free but a biased
random walk with asymetric jumping probabilities $p\left(\overrightarrow{r}_{i},\overrightarrow{r}_{i}^{\prime},t\right)\neq p\left(\overrightarrow{r}_{i}^{\prime},\overrightarrow{r}_{i},t\right)$,
which on average can be described by the well-known Smoluchowski equation
for the probability distribution\begin{equation}
\partial_{t}P\left(t,\vec{r}\right)=\left(\overrightarrow{f}\left(t,\vec{r}\right)\cdot\overrightarrow{\nabla}\right)P\left(t,\vec{r}\right)+D\left(t,\vec{r}\right)\triangle P\left(t,\vec{r}\right),\label{EQ3}\end{equation}
where the drift force term $\overrightarrow{f}\left(t,\vec{r}\right)$
and the diffusion coefficient $D\left(t,\vec{r}\right)$ are respectively
the first and second moments of the transition probabilities $p$.

The numerical simulation of the resulting {}``drifted\textquotedblright{}
DLA model yields pictures that are in a better agreement with experimental
results. One can go beyond this subjective visual estimate and employ
the fractal dimension as a quantitative characteristic of the dendritic
forms considered which allows for a meaningful and objective comparison
between the model and the experiments. We show that, even though the
Hausdorff-Minkowski dimension may be used for our purpose, the so-called
external fractal dimension is much more adapted due to its great sensitivity
to the intensity of the stress field. Confronting our model to experimental
results thanks to this tool, we find excellent agreement for antimony-cluster
structures. Our model, however, suggests that a repulsive force is
at work in the case of silver clusters, which is incompatible with
an elastic field.

The paper is structured as follows. In Sec. II, we present the drifted
DLA model in detail, provide simulations and discuss the influence
of the stress field on the morphologies obtained as well as their
visual agreement with experimentally observed structures. Sec. III
is devoted to the characterization of the ramified structures obtained,
either experimentally or numerically, through their fractal dimension.
In Sec. IV, this quantitative tool is used to discuss the agreement
between our model and some experimental results published in the literature.
We conclude in Sec. V and suggest some possible perspectives of our
work.

\section{The biased random walk model and the emerging shapes of cluster aggregations}

\begin{figure}[ptb]
\begin{centering}
\includegraphics[width=8cm]{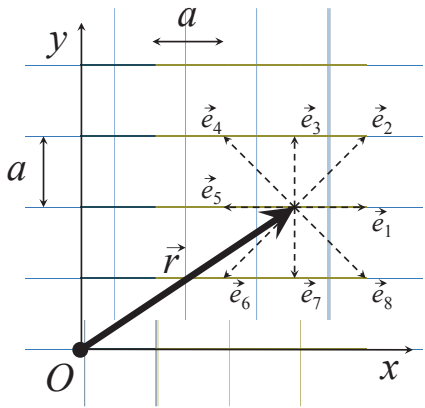} 
\par\end{centering}

\caption{The square lattice in the eight-connectivity model: each site is characterized
by its position vector $\vec{r}$ and connected to its eight neighbours
$\left(\vec{r}+\protect\overrightarrow{e_{i}}\right)_{i=1,\ldots,8}$.}

\label{Fig1} %
\end{figure}

We start by presenting some technical details of the numerical aproach
employed for the cluster aggregation modelling and the results of
simulations follow. The surface of the substrate is considered as
a two-dimensional square lattice of step $a$. Each site can store
one cluster at most, and its position is characterized by a vector
$\vec{r}$ (see Fig. \ref{Fig1}).

At the beginning of a deposition experiment, a first cluster is dropped
onto the surface and diffuses freely on the substrate. At each time
step of its random walk, this cluster can thus jump from the site
$\vec{r}$ it occupies to one of its eight closest neighbours of respective
positions $\left(\vec{r}+\vec{e}_{i}\right)$, with the same (isotropic)
probability $p_{iso}=\frac{1}{8}$. At some point of its diffusive
exploration of the surface, this cluster reaches a default of the
substrate where it stops. The corresponding site is arbitrarily chosen
as the origin $O$ of the coordinates -- its position vector is therefore
$\vec{s}_{1}=\vec{0}$. This first cluster is now ready to play the
role of a seed for a ramified fractal-like structure.

A second cluster is then dropped onto the surface at a random site
-- not {}``too far'' from the origin, otherwise it would take it
too long to reach the seed by diffusion, and diffuses on the substrate
now subject to the force implied by the interaction with the first
cluster fixed at the origin. Assuming this force is proportional to
the elastic deformation of the substrate and therefore scales like
$1/r$ according to Hooke's law \cite{landau}, it can be viewed as
deriving from the inter-cluster interaction potential $v\left(r\right)=\alpha\ln\left(r/r_{0}\right)$,
where the length parameter $r_{0}$ and the energy parameter $\alpha$
characterize the potential range and intensity, respectively. The
relevant force $\vec{F}\left(\vec{r}\right)=-\vec{\nabla}_{\vec{r}}v\left(\vec{r}\right)=-\alpha\vec{r}/r^{2}$
acting on the moving cluster affects the transition probabilities
which now take the form \begin{equation}
p\left(\vec{r},\vec{r}+\vec{e}_{i}\right)\simeq\frac{1}{8}-\frac{\beta}{\left\Vert \vec{e}_{i}\right\Vert }\left(\vec{F}\left(\vec{r}\right)\cdot\vec{e}_{i}\right),\label{EQ4}\end{equation}
 with $\beta=1/k_{B}T$, as it follows from the Smoluchowski equation.
For this expression to be valid, the anisotropic term $-\beta\vec{F}_{1}\left(\vec{r}\right)\cdot\frac{\vec{e}_{i}}{\left\Vert \vec{e}_{i}\right\Vert }$
must remain small compared to the isotropic probability $p_{iso}=\frac{1}{8}$,
which ensures that the cluster motion never reaches the balistic regime
but keeps its diffusive character. The second cluster stops its motion
at a point $\vec{s}_{2}$, as soon as it {}``touches'' the seed
located at the origin, or in other terms, once it occupies one of
the eight sites closest to the origin.

A third cluster is then dropped onto the substrate at a random starting
position -- which, again, should not be chosen too far from the two-cluster
island already formed. This cluster also diffuses on the surface of
the substrate. Its diffusive motion is however biased by an elastic
force due to the two-cluster island located around the origin. This
force is assumed to derive from the sum $V\left(\vec{r}\right)=\alpha\ln\left(\left\Vert \vec{r}-\vec{s}_{1}\right\Vert /r_{0}\right)+\alpha\ln\left(\left\Vert \vec{r}-\vec{s}_{2}\right\Vert /r_{0}\right)$
of the potentials induced by the first two clusters. The jump probability
$p\left(\vec{r},\vec{r}+\vec{e}_{i}\right)$ is still given by Eq.(\ref{EQ4})
but now with $\vec{F}\left(\vec{r}\right)\equiv-\vec{\nabla}_{\vec{r}}V\left(\vec{r}\right)$.
The biased diffusion stops when the third cluster visits for the first
time one of the neighboring sites of the first two aggregated clusters,
of position $\vec{s}_{3}$. The general case of the $N^{\mathrm{th}}$
deposited cluster follows by analogy.

\begin{widetext}

\begin{figure}[ptb]
\begin{centering}
\includegraphics[width=12cm]{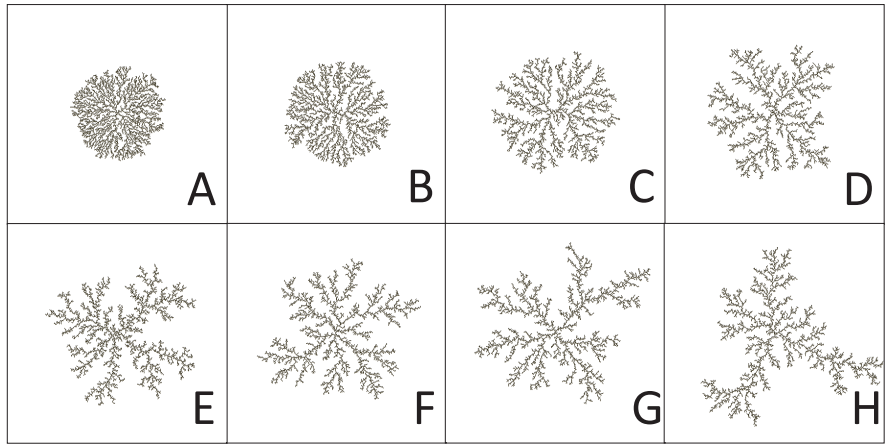}
\par\end{centering}

\caption{Simulations of our ramified structures obtained in our drifted DLA
model by the deposition $6400$ clusters for different values of the
dimensionless parameter $\kappa$: (A) $\kappa=3,3\times10^{-3}$,
(B) $\kappa=1\times10^{-3}$, (C) $\kappa=3,3\times10^{-4}$, (D)
$\kappa=1\times10^{-4}$, (E) $\kappa=3,3\times10^{-5}$, (F) $\kappa=0$
corresponding to the standard DLA model, (G) $\kappa=-1\times10^{-5}$
and (H) $\kappa=-3,3\times10^{-5}$.}
\label{Fig2} %
\end{figure}
\end{widetext}

Note that in this model, the clusters are successively dropped onto
the surface, as it is usually assumed in the DLA approach. Though
in a real experiment many clusters diffuse over the surface simultaneously,
the interaction among them is negligible as compared to the potential
induced by the clusters already agregated in a ramified structure.
Also note that, in our model, we never make use of any specific property
of the interaction mediated by elastic forces. The nature of the interaction
might therefore be of a different origin, and the only requirement
is that the interaction potential satisfies \begin{equation}
\triangle V\left(\vec{r}\right)=\alpha c\left(\vec{r}\right),\label{EQ5}\end{equation}
 where $c\left(\vec{r}\right)=1$ for sites occupied by aggregated
clusters, and $c\left(\vec{r}\right)=0$ otherwise. As a consequence,
it is also possible to make simulations for repulsive potentials,
\emph{i.e.} with $\alpha<0$, which do not correspond to elastic deformations.

In Fig. \ref{Fig2} we present a few simulated pictures obtained through
the aggregation of $N=6400$ clusters whose diffusion was biased by
the potential Eq.(\ref{EQ5}) for different values of the dimensionless
parameter $\kappa\equiv\alpha\beta$. As could be expected, for attractive
forces $\kappa>0$ (Fig. \ref{Fig2} A-E), the structures have more
compact of the morphologies than those obtained in the DLA model $\kappa=0$
(Fig. \ref{Fig2} F). In the repulsive regime $\kappa<0$ (Fig. \ref{Fig2}
G,H), the \textquotedbl{}arms\textquotedbl{} of the ramified structures,
on the contrary tend to repel each other, thus giving rise to less
compact structures. The transition from sparse to compact objects
resembles one that has already been observed and modeled in the case
of electro-deposition experiments \cite{myochain2002,myochain2004},
although not for the case of self-consistent interaction but for a
fixed external potential.

\begin{widetext}

\begin{figure}[ptb]
\begin{centering}
\includegraphics[width=10cm]{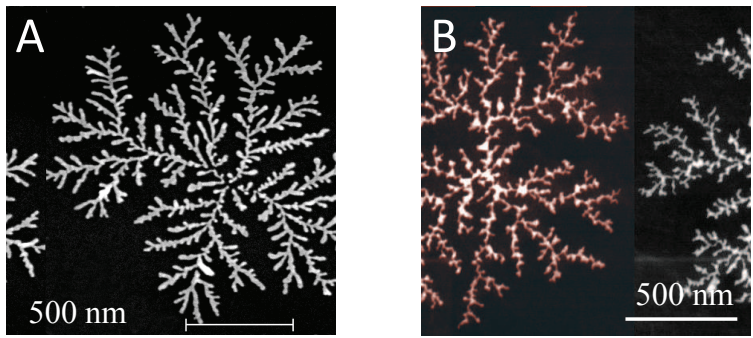} 
\par\end{centering}

\caption{Experimental pictures of single dendrites obtained through the deposition
of (A) silver $Ag_{100}$ \cite{brechi,lando} and (B) antimony $Sb_{500}$
\cite{yoon} atomic clusters onto graphite HOPG.}
\label{Fig6}%
\end{figure}
\end{widetext}

In Fig. \ref{Fig6} for comparison we show dendritic islands experimentally
obtained through depositing (A) silver Ag$_{100}$ \cite{brechi,lando}
and (B) antimony $Sb_{500}$ atomic clusters \cite{yoon}. At first
glance, \ref{Fig6} B (antimony clusters) seem to be well described
by the same kind of structures as in Fig. \ref{Fig2} D, corresponding
to our drifted DLA model, with a moderate attractive force, whereas
\ref{Fig6} A (silver clusters) seem to correspond to none of the
structures displayed in Fig. \ref{Fig2}, \emph{i.e. }it is satisfactorily
described by neither the DLA (Fig. \ref{Fig2} F) nore the drifted
DLA models (Fig. \ref{Fig2} A-E,G,H). One, however, needs a more
formal tool for quantification of this observation, that would enable
one to say how much the simulated structures match the experimental
data. In the next section, we introduce such quantities, the regular
and the external fractal dimensions. The latter presents the advantage
of being particularly sensitive to the elastic force at the core of
our model.

\section{Characterization method, discussion and results}

The fractal dimension appears as the most natural mathematical tool
to characterize ramified structures such as the dendritic islands
formed by aggregated atomic clusters, and was intensively used in
previous works \cite{hwang}. There, however, exist different operational
definitions of this quantity in the litterature which are not equivalent.
For instance, the methods based on the radius of gyration \cite{menshutin}
or the auto-correlation function \cite{witten} as well as multifractal
analysis techniques \cite{menshutinmultifrac} appear very sensitive
to the specific features of the structure under consideration. For
different DLA simulated morphologies, the fractal dimensions calculated
through these methods already show a very large dispersion, typically
ranging from $1.67$ to $1.72$. Because of this dispersion, these
methods cannot be expected to precisely distinguish simulations obtained
through different types of models.

\begin{figure}[ptb]
\begin{centering}
\includegraphics[width=10cm]{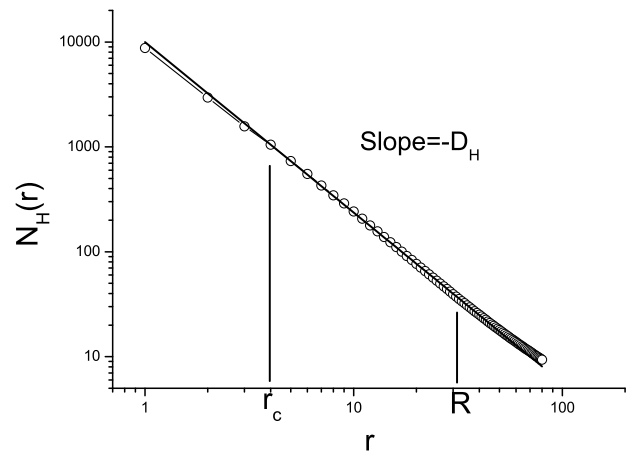} 
\par\end{centering}

\caption{The number of disks $N_{H}\left(r\right)$ covering a ramified simulated
structure as a function of the probe disks' radius $r$, in logarithmic
scales. Self-similarity of the fractal-like dendritic structure can
be checked on the range $\left[r_{c},R\right]$. }
\label{Fig3}%
\end{figure}

What we are looking for is, on the contrary, a quantity able to universally
characterize all the morphologies simulated by a given model. The
fractal dimension $D_{H}$ calculated according to the Hausdorff-Minkowski
definition precisely has this property. As shown in \cite{hann},
it takes the value $D_{H}=1.60$ for any DLA simulated structure with
the negligible dispersion $\pm0.01$, and we numerically checked the
same property for the drifted DLA model. To compute $D_{H}$ one covers
the object with disks of radius $r$, and measures the area $I_{H}(r)$
of the object thus formed. The number $N_{H}(r)$ of these disks is
then simply given by $N_{H}(r)=\frac{I_{H}(r)}{\pi r^{2}}$. The fractal
dimension $D_{H}$ is, by definition, determined by the expected asymptotic
behaviour $N_{H}(r)\sim r^{-D_{H}}$ when $r\rightarrow0$. Practically,
the size of the probe disks should, however, remain larger than the
typical size $r_{c}$ of a cluster in order to avoid spurious and
meaningless discretization effects. It must also not be too large,
that is, typically, smaller than the radius $R$ of the whole aggregate
-- when $r$ exceeds $R$ only one disk is necessary to cover the
structure and therefore $N_{H}(r)=1$. Finally, the quantity $-D_{H}=\underset{r\rightarrow0,r\gg r_{c}}{\lim}\frac{\ln N_{H}(r)}{\ln r}$
appears as the slope of the approximately linear function $\ln N_{H}(r)=f\left(\ln r\right)$
on the intermediate range $\left[r_{c},R\right]$ (see Fig.\ref{Fig3}).

\begin{widetext}

\begin{figure}[ptb]
\begin{centering}
\includegraphics[width=12cm]{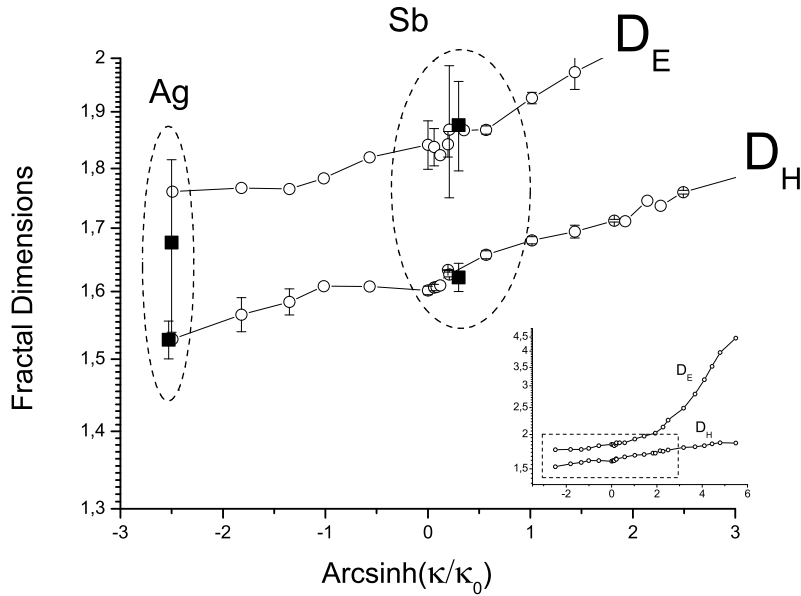}
\par\end{centering}

\caption{The\textbf{\ inset} presents the behaviours of $D_{H}$ (lower curve)
and $D_{E}$ (upper curve) of the morphologies obtained by the numerical
simulation of our drifted DLA model, as functions of the parameter
$\mathrm{arcsinh}\left(\kappa/\kappa_{0}\right)$ characterizing the
intensity of the elastic field induced by the growing island ($\kappa_{0}=0.17\times10^{-4}$).
The \textbf{main plot} is a zoom of the inset on the area contained
in the dotted box. Two couples of points have been reported, corresponding
to the values of $D_{H}$ and $D_{E}$, extracted from a set of experimental
pictures obtained by the deposition of silver $Ag_{100}$ \cite{brechi,lando}
and antimony $Sb_{500}$ atomic clusters \cite{yoon} onto graphite
HOPG. The value of the parameter $\kappa$ is adjusted so as to minimize
$\left|D_{H}^{\mathrm{th}}\left(\kappa\right)-D_{H}^{\mathrm{exp}}\right|+\left|D_{E}^{\mathrm{th}}\left(\kappa\right)-D_{E}^{\mathrm{exp}}\right|$
in both cases. The points corresponding to antimony can be reported
on the theoretical plot with excellent agreement and correspond to
$\kappa_{\mathrm{Sb}}\simeq5\times10^{-6}$ (small attractive potential),
while our model seems to fail to correctly describe the experimental
values for silver. }
\label{Fig5} %
\end{figure}
\end{widetext}

The lower curve in Fig. \ref{Fig5} shows the fractal dimension $D_{H}$
of simulated structures as a function of $\kappa$. As expected, the
dimension $D_{H}$ grows with the intensity of the force, which is
consistent with the visible increasing compactness of the forms numerically
generated. The variations of $D_{H}$, however, remain very moderate:
quantifying the effects of the elastic field through $D_{H}$ will
therefore not constitute a very sensitive method.

To remedy this problem, we consider an aternative possible characteristic
of a ramified fractal-like morphology, the so-called external fractal
dimension $D_{E}$, which turns out to be much more sensitive to the
variations of $\kappa$. To our knowledge, this fractal dimension
has never been applied in the context of cluster aggregated structures.
By definition, $D_{E}$ is the fractal dimension of the periphery
of the object. To be more explicit, one computes $D_{E}$ by measuring
the area $I_{E}(r)$ swept by a disk of radius $r$, sliding along
the border of the structure. Intuitively, in a ramified object, as
$r$ decreases, more and more structures appear. This results in a
non trivial behaviour of $I_{E}(r)$, and therefore of $N_{E}(r)=\frac{I_{E}(r)}{\pi r^{2}}$,
the number of disks necessary for covering the whole area $I_{E}(r)$.
For a rigorously fractal object $N_{E}(r)\varpropto r^{-D_{E}}$.
The same scaling is found in morphologies showing self-similarity
on a given range of radii, as the structures generated by our simulations.

The behaviours of $D_{E}$ and $D_{H}$ can be compared in the inset
of Fig. \ref{Fig5}. In particular, it is quite apparent that $D_{E}$
is much more sensitive to $\kappa$ than $D_{H}$ and is therefore
a better suited quantity for characterizing the modifications imposed
to the aggregated structures by the elastic field.

\section{Comparison with experiments}

Let us now confront our model to some experimental data available
in the literature. We first consider experimental pictures of dendritic
structures obtained through depositing Sb$_{500}$ antimony atomic
clusters on a graphite HOPG substrate \cite{yoon}, such as Fig. \ref{Fig6}
B. Following the same procedures as described above, we extract the
Hausdorff and external fractal dimensions of the set of $\sim100$
pictures we have, and get the averaged values $D_{H}^{\mathrm{exp}}\left(\mathrm{Sb}\right)\simeq1.622\pm0.022$
and $D_{E}^{\mathrm{exp}}\left(\mathrm{Sb}\right)\simeq1.876\pm0.080$.
If our model correctly describes the aggregation of antimony clusters,
there must exist a specific value of the parameter $\kappa$, denoted
by $\kappa_{\mathrm{Sb}}$, such that drifted-DLA simulations run
with this specific parameter $\kappa_{\mathrm{Sb}}$ yield values
for $D_{H}$ and $D_{E}$ which coincide with the experimental ones.
In other terms, there should exist a value $\kappa_{\mathrm{Sb}}$
such that $D_{H}^{\mathrm{exp}}\left(\mathrm{Sb}\right)\simeq D_{H}^{\mathrm{sim}}\left(\kappa_{\mathrm{Sb}}\right)$
and $D_{H}^{\mathrm{exp}}\left(\mathrm{Sb}\right)\simeq D_{E}^{\mathrm{sim}}\left(\kappa_{\mathrm{Sb}}\right)$
are simultaneously checked -- here, $D_{H,E}^{\mathrm{sim}}\left(\kappa\right)$
denotes the Hausdorff/external fractal dimension of the structures
obtained by drifted-DLA simulations with the parameter $\kappa$.
To identify such a value, one simply tries to numerically minimize
to zero the quantity $\left\vert D_{H}^{\mathrm{sim}}\left(\kappa\right)-D_{H}^{\mathrm{exp}}\left(\mathrm{Sb}\right)\right\vert +\left\vert D_{E}^{\mathrm{sim}}\left(\kappa\right)-D_{E}^{\mathrm{exp}}\left(\mathrm{Sb}\right)\right\vert $.
It turns out that the minimization is successful and yields the value
$\kappa_{\mathrm{Sb}}\simeq5\times10^{-6}$. The two experimental
values $D_{H,E}^{\mathrm{exp}}\left(\mathrm{Sb}\right)$ are reported
on Fig. \ref{Fig5}: the agreement is excellent indeed between model
and experiment, which suggests that the physics of aggregation of
antimony clusters is actually correctly described by our model and
involves a small attractive elastic potential. It is also interesting
to note that $\kappa_{\mathrm{Sb}}\simeq5\times10^{-6}$ is precisely
the value corresponding to the picture Fig.\ref{Fig2}D: this means
that the experimental picture Fig.\ref{Fig6}B is indeed better described
by a structure of the kind of Fig.\ref{Fig2}D than by a pure DLA
morphology Fig.\ref{Fig2}F. The visual impression we had at the end
of Sec. II is therefore completely confirmed by our quantitative analysis.

We now turn to pictures obtained by depositing Ag$_{100}$ silver
atomic clusters \cite{brechi,lando}. The experimental values we extract
are $D_{H}^{\mathrm{exp}}\left(\mathrm{Ag}\right)\simeq1.528\pm0.028$
and $D_{E}^{\mathrm{exp}}\left(\mathrm{Ag}\right)\simeq1.677\pm0.138$.
Note that the error bars, especially on $D_{E}^{\mathrm{exp}}\left(\mathrm{Ag}\right)$,
are too important to make any firm conclusion. This dispersion follows
from the small size of the dendrites observed on experimental pictures,
which substantially truncates the range $\left[r_{c},R\right]$ on
which self-similarity can be observed. The fractal dimension therefore
cannot be extracted with a sufficiently high fidelity. If, however,
we apply the same procedure as for antimony, we observe that there
exists no value of $\kappa$ minimizing to zero the quantity $\left|D_{H}^{\mathrm{sim}}\left(\kappa\right)-D_{H}^{\mathrm{exp}}\left(\mathrm{Ag}\right)\right|+\left|D_{E}^{\mathrm{sim}}\left(\kappa\right)-D_{E}^{\mathrm{exp}}\left(\mathrm{Ag}\right)\right|$.
The experimental values $D_{H,E}^{\mathrm{exp}}\left(\mathrm{Ag}\right)$
reported on Fig. \ref{Fig5} correspond to the value $\kappa_{\mathrm{Ag}}\simeq-10^{-4}$
of the parameter $\kappa$ providing the best compromise, though clearly
far from being satisfactory. Note that $\kappa_{\mathrm{Ag}}<0$ does
not correspond to an elastic (attractive) field, the physics of aggregation
therefore does not seem governed by the same mechanisms as for Sb.
More investigations need be taken to better understand how to account
for this specific behaviour.

\section{Conclusion}

We have presented a new class of biased diffusion models for the aggregation
of atomic clusters deposited on a plane substrate. This approach can
be viewed as a development generalizing the traditionnally used DLA
model, that we completed by introducing an elastic force originating
from the substrate deformation by the growing structure. The diffusion
equation which governs the cluster density dynamics is therefore transformed
into the Smoluchowski equation. Numerical simulations of our model
yield ramified structures, whose compactness is seen to increase with
the intensity of the deformation force.

To quantitatively compare the results of our simulations with experimental
data, we investigated Hausdorff and external fractal dimensions, and
showed that the latter is better suited for characterizing the modifications
of the structures obtained in our model due to the elastic field.

Finally, we compared our simulations to already published experimental
results, on silver and antimony clusters. Whereas our model describes
antimony-cluster-aggregated structures very well, it is not adapted
to silver. This suggests that the dynamics of aggregation is dominated
by different processes, depending on the atomic nature of the clusters
considered. Identifying what these physical processes are, and their
exact influence on the form of the morphologies of the corresponding
ramified structures is an interesting perspective of the present work.
Another possible investigation would consist to apply the same kind
of analysis, involving the fractal dimension, to physically completely
different domains, such as electro-deposition where DLA approach clearly
fails \cite{myochain2004}.
\begin{acknowledgments}
The authors thank F. Prats for fruitful discussions concerning numerical
simulations. \end{acknowledgments}

\end{document}